\documentclass{amsart}
\usepackage{latexsym,amsfonts,amsmath,amssymb}
\newcommand{\Reals}{{\mathbb{R}}}
\newcommand{\Cmplx}{{\mathbb{C}}}

\newcommand{\IR}{{\mathbb{R}}}
\newcommand{\IC}{{\mathbb{C}}}

\newcommand{\ID}{{\mathbb{D}}}

\newcommand{\Disk}{{\mathbb{D}}}
\newcommand{\Exp}{{\mathbb{E}}}

\newcommand{\RR}{{{}_{R}}}

\newcommand{\g}{{\mathfrak{g}}}
\newcommand{\A}{{\mathfrak{a}}}
\newcommand{\U}{{\mathbb{U}}}
\newcommand{\LL}{{\mathfrak{l}}}
\newcommand{\LLL}{{\mathbb{L}}}

\newcommand{\C}{{\mathcal{C}}}

\newcommand{\al}{{\alpha}}

\let\det=\undefined\DeclareMathOperator*{\det}{det}
\let\Re=\undefined\DeclareMathOperator*{\Re}{Re}
\let\Im=\undefined\DeclareMathOperator*{\Im}{Im}

\DeclareMathOperator*{\tr}{tr} \DeclareMathOperator*{\diag}{diag}

\newtheorem{theorem}{Theorem}%[section]
\newtheorem{prop}[theorem]{Proposition}

\theoremstyle{definition}

\newcounter{smalllist}

\let\llldots=\ldots
\def\ldots{\llldots{}}

\numberwithin{equation}{section}

\begin{document}

\title[CMV matrices: a survey]{CMV matrices in random matrix theory and integrable systems: a survey}
\author{Irina Nenciu}
\address{Irina Nenciu\\
         Courant Institute\\
         251 Mercer St.\\
         New York, NY 10012}
\email{nenciu@cims.nyu.edu}
%\date{\today}

\begin{abstract}
We present a survey of recent results concerning a remarkable class of unitary matrices,
the CMV matrices. We are particularly interested in the role they play in the theory
of random matrices and integrable systems. Throughout the paper we also
emphasize the analogies and connections to Jacobi matrices.
\end{abstract}

\maketitle

%%%%%%%%%%%%%%%%%%%%%%%%%%%%%%%%%%%%%%%%%%%%%%%%%%%%%%%%%%%%%%%%%%%%%%%%%%%%%%%%%%%%%
\section{Introduction}
%%%%%%%%%%%%%%%%%%%%%%%%%%%%%%%%%%%%%%%%%%%%%%%%%%%%%%%%%%%%%%%%%%%%%%%%%%%%%%%%%%%%%

From many points of view, Jacobi matrices play a central role among all Hermitian matrices.
We wish to discuss a class of unitary matrices recently introduced by Cantero, Moral, and Vel\'azquez,
\cite{CMV}, which play a similar role among unitary matrices. While this analogy may be extended in many directions,
we investigate here their occurrence in the theory of random matrices and integrable systems.

By Jacobi matrix we mean an $n\times n$ tri-diagonal matrix
\begin{equation}\label{Jmat}
J=\begin{bmatrix}
 b_1  &  a_1 &        & \\
 a_1  &  b_2 & \ddots & \\
      &\ddots& \ddots & a_{n-1}\\
      &      & a_{n-1}&  b_n
\end{bmatrix}
\end{equation}
with $a_j>0$, $b_j\in\Reals$.

Given coefficients $\alpha_0,\ldots,\alpha_{n-2}\in\Disk$, the unit disk in $\IC$, and
$\alpha_{n-1}\in S^1$, let $\rho_k=\sqrt{1-|\alpha_k|^2}$, and
define $2\times 2$ matrices
$$
\Xi_k = \begin{bmatrix} \bar\alpha_k & \rho_k \\ \rho_k & -\alpha_k
\end{bmatrix}
$$
for $0\leq k\leq n-2$, while $\Xi_{-1}=[1]$ and
$\Xi_{n-1}=[\bar\alpha_{n-1}]$ are $1 \times1$ matrices. From these,
form the $n\times n$  block-diagonal matrices
$$
\mathcal{L}=\diag\bigl(\Xi_0   ,\Xi_2,\Xi_4,\ldots\bigr)
\quad\text{and}\quad
\mathcal{M}=\diag\bigl(\Xi_{-1},\Xi_1,\Xi_3,\ldots\bigr).
$$
The \emph{CMV matrix} associated to the coefficients
$\alpha_0,\ldots,\alpha_{n-1}$ is $\C=\mathcal{LM}$.
It is a 5-diagonal unitary matrix, given by:
\begin{equation}
\mathcal{C}=
\left(%
\begin{array}{ccccccc}
  \bar\alpha_0 & \rho_0\bar\alpha_1    & \rho_0\rho_1    & 0 & 0 & 0 & \ldots \\
  \rho_0       & -\alpha_0\bar\alpha_1 & -\alpha_0\rho_1 & 0 & 0 & 0 & \ldots \\
  0 & \rho_1\bar\alpha_2 & -\alpha_1\bar\alpha_2 & \rho_2\bar\alpha_3 & \rho_2\rho_3 & 0 & \ldots \\
  0 & \rho_1\rho_2 & -\alpha_1\rho_2 & -\alpha_2\bar\alpha_3 & -\alpha_2\rho_3 & 0 & \ldots \\
  0 & 0 & 0 & \rho_3\bar\alpha_4 & -\alpha_3\bar\alpha_4 & \rho_4\bar\alpha_5 & \ldots \\
  0 & 0 & 0 & \rho_3\rho_4 & -\alpha_3\rho_4 & -\alpha_4\bar\alpha_5 & \ldots \\
  \ldots & \ldots & \ldots & \ldots & \ldots & \ldots & \ldots \\
\end{array}%
\right).
\end{equation}

Following \cite{Simon1}, we will refer to the numbers $\alpha_k$ as Verblunsky coefficients.
A related system of matrices was discovered independently by Tao and Thiele, \cite{TaoThiele},
in connection with the non-linear Fourier transform.  Their matrices are bi-infinite and
correspond to setting odd-indexed Verblunsky coefficients to zero.  (This has the effect of
doubling the spectrum; cf. \cite[p. 84]{Simon1}.)

Both Jacobi and CMV matrices arose in the study of orthogonal polynomials, so it is
natural that we begin there. We will first describe the relation of
orthogonal polynomials to Jacobi matrices and then explain the
connection to CMV matrices.

Given a probability measure $d\nu$ supported on a finite subset of
$\Reals$, say of cardinality $n$, we can apply the Gram--Schmidt
procedure to $\{1,x,x^2,\ldots,x^{n-1}\}$ and so obtain an
orthonormal basis for $L^2(d\nu)$ consisting of
polynomials, $\{p_j(x):j=0,\ldots,{n-1}\}$, with positive leading
coefficient. In this basis, the linear transformation $f(x)\mapsto
xf(x)$ is represented by a Jacobi matrix.  An equivalent statement
is that the orthonormal polynomials obey a three-term recurrence:
$$
xp_{j}(x) = a_{j} p_{j+1}(x) + b_{j}p_j(x) + a_{j-1} p_{j-1}(x)
$$
where $a_{-1}=0$ and $p_{n}\equiv 0$.

We have just shown how measures on $\Reals$ lead to Jacobi matrices;
in fact, there is a one-to-one correspondence between them.  Given a
Jacobi matrix, $J$, let $d\nu$ be the spectral measure associated to
$J$ and the vector $e_1=[1,0,\ldots,0]^{T}$.  Then $J$ represents
$f(x)\mapsto xf(x)$ in the basis of orthonormal polynomials associated
to $d\nu$.

To explain the origin of CMV matrices, it is necessary to
delve a little into the theory of orthogonal polynomials on the unit
circle.  For a more complete description of what follows, the reader
should turn to \cite{Simon1}.

Given a finitely-supported probability measure $d\mu$ on $S^1$, the
unit circle in $\Cmplx$, we can construct an orthonormal system of
polynomials, $\phi_k$, by applying the Gram--Schmidt procedure to
$\{1,z,\ldots\}$.  These obey a recurrence relation; however, to
simplify the formulae, we will present the relation for the monic
orthogonal polynomials $\Phi_k(z)$
\begin{align}
\Phi_{k+1}(z)   &= z\Phi_k(z)   - \bar\alpha_k \Phi_k^*(z).
\label{PhiRec}
\end{align}
Here $\Phi_k^*$ denotes the reversed polynomial:
\begin{equation}\label{rev}
\Phi_k(z) = \sum_{l=0}^k c_l z^l \quad \Rightarrow \quad \Phi_k^*(z)
= \sum_{l=0}^k \bar{c}_{k-l} z^l=z^k\overline{\Phi_k\left(\tfrac{1}{\bar z}\right)},
\end{equation}
and $\alpha_k$ are recurrence coefficients. They are in fact the same as the
Verblunsky coefficients which appeared above, in the definition of the CMV matrix.
When $d\mu$ is supported at exactly $n$ points, $\alpha_k\in\Disk$
for $0\leq k\leq {n-2}$, while $\alpha_{n-1}$ is a unimodular complex
number.

The Verblunsky coefficients completely describe the measure $d\mu$:
There is a 1-to-1 correspondence between probability measures on the unit circle
supported at $n$ points and Verblunsky coefficients $(\alpha_0,\ldots,\alpha_{n-1})$
with $\alpha_{k}\in\ID$ for $0\leq k\leq n-2$ and $\alpha_{n-1}\in S^1$.

From the discussion of Jacobi matrices, it would be natural to consider the matrix representation
of $f(z)\mapsto zf(z)$ in $L^2(d\mu)$ with respect to the basis of orthonormal polynomials. This is
\textit{not} a CMV matrix; rather it is a Hessenberg matrix which Simon, \cite{Simon1}, has dubbed a GGT matrix, from the
initials of Geronimus, Gragg, and Teplyaev.  Perhaps the most striking difference from a CMV (or
Jacobi) matrix is that a GGT matrix is very far from sparse---generically, all entries above and
including the sub-diagonal are non-zero.

Cantero, Moral, and Vel\'azquez had the simple and ingenious idea of applying the Gram--Schmidt
procedure to $\{1,z,z^{-1},z^2,z^{-2},\ldots\}$ rather than $\{1,z,\ldots\}$. The resulting
functions, $\chi_k(z)$ ($0\leq k\leq n-1$), form a basis and are easily expressible in terms of the orthonormal polynomials.
In this basis, the map $f(z)\mapsto zf(z)$ is represented in an especially simple form:

\begin{theorem}[Cantero, Moral, Vel\'azquez, 2003]\label{T:2a}
In the orthonormal basis $\{\chi_k(z)\}$ of $L^2(d\mu)$, the operator $f(z)\mapsto zf(z)$ is
represented by the CMV matrix $\C$ associated to the Verblunsky coefficients of the measure $d\mu$.
\end{theorem}

The $\C=\mathcal{LM}$ factorization presented above originates as follows:
Let us write $x_k$, $0\leq k\leq n-1$, for the orthonormal basis constructed by applying the Gram--Schmidt
procedure to $\{1,z^{-1},z,z^{-2},z^2,\ldots\}$.  Then the matrix elements of $\mathcal{L}$ and
$\mathcal{M}$ are given by
$$
\mathcal{L}_{j+1,k+1}=\langle \chi_j(z) |z  x_k(z) \rangle, \qquad
\mathcal{M}_{j+1,k+1}=\langle x_j(z) | \chi_k(z) \rangle.
$$
See \cite{Simon1} for further discussion.

The measure $d\mu$ can be reconstructed from $\C$ in a manner analogous to the Jacobi case.
Let $d\mu$ be the spectral measure associated to a CMV matrix, $\C$,
and the vector $e_1$.  Then $\C$ is the CMV matrix associated to the
measure $d\mu$.

Proofs of these statements can be found in \cite{CMV} or \cite{Simon1}.

An interesting special case is when the measure $d\mu$ is symmetric with respect to complex conjugation.
This occurs if and only if all Verblunsky coefficients are real.  It is a famous observation of
Szeg\H{o} (see \cite[\S11.5]{Szego}) that the polynomials orthogonal
with respect to this measure are intimately related to the
polynomials orthogonal with respect to the measure $d\nu$ on
$[-2,2]$ defined by
\begin{equation}\label{nuDefn}
  \int_{S^1} f(z+z^{-1}) \, d\mu(z)  =  \int_{-2}^2 f(x) \,d\nu(x).
\end{equation}
Moreover, the recurrence coefficients for
these measures are related by the Geronimus relations:
\begin{equation}\label{AB:Geron}
\begin{aligned}
b_{k+1}   &= (1-\alpha_{2k-1})\alpha_{2k} -
(1+\alpha_{2k-1})\alpha_{2k-2}           \\
a_{k+1}   &= \big\{
(1-\alpha_{2k-1})(1-\alpha_{2k}^2)(1+\alpha_{2k+1}) \big\}^{1/2}.
\end{aligned}
\end{equation}
The classical proof of \eqref{AB:Geron} relies on the relation between the orthogonal polynomials
associated to the two measures. An alternate proof was given by Rowan Killip and the author,
\cite[Proposition~B.3]{KN}, who derive \eqref{AB:Geron} from an explicit relation between the CMV and
Jacobi matrices of the two measures $d\mu$ and $d\nu$, respectively.

%%%%%%%%%%%%%%%%%%%%%%%%%%%%%%%%%%%%%%%%%%%%%%%%%%%%%%%%%%%%%%%%%%%%%%%%%%%%%%%%%%%%%
\section{Circular $\beta$-ensembles}
%%%%%%%%%%%%%%%%%%%%%%%%%%%%%%%%%%%%%%%%%%%%%%%%%%%%%%%%%%%%%%%%%%%%%%%%%%%%%%%%%%%%%

In 1962, Dyson \cite{Dyson} introduced three ensembles of random
unitary matrices with a view to simplifying the study of energy
level behavior in complex quantum systems. Earlier work in this
direction, pioneered by Wigner, focused on ensembles of Hermitian
matrices. In both of these situations, the eigenvalue distributions coincide
with the Gibbs distribution for particles of Coulomb gas at inverse temperature
$\beta=1,2,4$ on the circle or the real line, respectively.
Dumitriu and Edelman, \cite{DumE}, constructed tri-diagonal matrix models for two
of the three standard examples of the Coulomb gas on the real line for \textit{any}
inverse temperature: the $\beta$-Hermite and the $\beta$-Laguerre. Rowan Killip and the author
obtained the analogous result on the circle, using CMV matrices.

One of the main features of all of these models is that they are defined in terms
of $O(n)$ independent random variables which appear in matrix entries
via very simple combinations. Moreover, the matrices are all sparse, having
$O(n)$ non-zero entries. Finally, these are not asymptotic models (see \cite{FR} and \cite{Lip}),
but exact ones. (Throughout this section, $n$ denotes the number of particles.)

Let us elaborate: The easiest way to obtain a normalizable Gibbs measure on the real
line is to add an external harmonic potential
$V(x)=\tfrac{1}{2}x^2$.  This gives rise to the probability measure
\begin{equation}\label{CGRL}
\Exp(f) \propto \int\!\! \cdots \!\! \int f(x_1,\ldots,x_n) \,
    \bigl|\Delta(x_1,\ldots,x_n)\bigr|^\beta  \prod_j e^{-V(x_j)}  \,
dx_1\cdots dx_n
\end{equation}
on $\Reals^n$, where, as usual, $\Delta$ denotes the Vandermonde determinant.
This is known as the Hermite ensemble, because of
its intimate connection to the orthogonal polynomials of the same
name, and when $\beta=1$, $2$, or $4$, arises as the eigenvalue
distribution in the classical Gaussian ensembles of random matrix
theory.  Dumitriu and Edelman showed that \eqref{CGRL} is the
distribution of eigenvalues for a symmetric tri-diagonal matrix with
independent entries (modulo symmetry).  The diagonal entries have
standard Gaussian distribution and the lower diagonal entries are
$2^{-1/2}$ times a $\chi$-distributed random variable with the
number of degrees of freedom equal to $\beta$ times the number of
the row.

The second example treated by Dumitriu and Edelman is the Laguerre
ensemble. In statistical circles, this is known as the Wishart
ensemble, special cases of which arise in the empirical
determination of the covariance matrix of a multivariate Gaussian
distribution. For this ensemble, one needs to modify the
distribution given in \eqref{CGRL} in two ways: each particle $x_j$
is confined to lie in $[0,\infty)$ and is subject to the external
potential $V(x)=-a\log(x)+x$, where $a>-1$ is a parameter. In
\cite{DumE} it is shown that if $B$ is a certain $n\times n$ matrix
with independent $\chi$-distributed entries on the main and
sub-diagonal (the number of degrees of freedom depends on $a$,
$\beta$, and the element in question) and zeros everywhere else,
then the eigenvalues of $L=BB^T$ follow this distribution.

The third standard ensemble on the real line is the Jacobi ensemble.
The distribution is as in \eqref{CGRL}, but now the
particles are confined to lie within $[-2,2]$ and are subject to the
external potential $V(x)=-a\log(2-x)-b\log(2+x)$, where $a,b>-1$ are
parameters.  This corresponds to the probability measure on
$[-2,2]^n$ that is proportional to
\begin{equation}\label{E:JE}
    \bigl|\Delta(x_1,\ldots,x_n)\bigr|^\beta  \prod_j \,
(2-x_j)^{a}(2+x_j)^{b}  \, dx_1\cdots dx_n.
\end{equation}
Dumitriu and Edelman did not give a matrix model for this ensemble,
listing it as an open problem.

On the unit circle, one does not need a confining potential in
order to define the Gibbs distribution for $n$ particles of the
Coulomb gas at inverse temperature $\beta$; it is given by
\begin{equation}\label{CGbeta}
\mathbb{E}^{\beta}_{n}(f)\propto \int\!\!\cdots\!\!\int\,
f(e^{i\theta_1},\ldots,e^{i\theta_n})|\Delta(e^{i\theta_1},\ldots,e^{i\theta_n}
)|^{\beta} \frac{d\theta_1}{2\pi}\cdots\frac{d\theta_n}{2\pi}
\end{equation}
for any symmetric function $f$. CMV matrices realize a random matrix model
with this eigenvalue distribution.

We say that a complex random variable, $X$, with values in the unit disk,
$\Disk$, is $\Theta_\nu$-distributed (for $\nu>1$) if
\begin{equation}\label{E:ThetaDefn}
\Exp\{f(X)\} = \tfrac{\nu-1}{2\pi} \int\!\!\!\int_\Disk f(z)
(1-|z|^2)^{(\nu-3)/2} \,d^2z.
\end{equation}
For $\nu\geq2$ an integer, this has the following geometric
interpretation: If $v$ is chosen from the unit sphere $S^\nu$ in
$\Reals^{\nu+1}$ at random according to the usual surface measure,
then $v_1+iv_2$ is $\Theta_\nu$-distributed. As a continuation of this geometric picture,
we shall say that $X$ is $\Theta_1$-distributed if it is uniformly distributed on the unit
circle in $\Cmplx$. With this definition, we have the following family of matrix models:

\begin{theorem}[Killip, Nenciu, 2004]\label{T:1}
Given $\beta>0$, let $\alpha_k\sim\Theta_{\beta(n-k-1)+1}$ be
independent random variables for $0\leq k\leq n-1$.
Then the CMV matrix $\C=\mathcal{LM}$ defined by these Verblunsky coefficients
gives a {\rm(}sparse{\rm)} matrix model for the
Coulomb gas at inverse temperature $\beta$. That is, its
eigenvalues are distributed according to the Gibbs distribution \eqref{CGbeta}.
\end{theorem}

Finally, we present a tri-diagonal matrix model for the Jacobi ensemble \eqref{E:JE}.
The independent parameters follow a beta distribution: A real-valued random variable $X$ is said
to be beta-distributed with parameters $s,t>0$, which we denote by $X\sim B(s,t)$, if
\begin{equation}\label{E:beta}
\Exp\{f(X)\} = \frac{2^{1-s-t}\Gamma(s+t)}{\Gamma(s)\Gamma(t)}
\int_{-1}^1 f(x) (1-x)^{s-1}(1+x)^{t-1} \, dx.
\end{equation}
Note that $B(\tfrac{\nu}{2},\tfrac{\nu}{2})$ is the distribution of
the first component of a random vector from the $\nu$-sphere.

\begin{theorem}[Killip, Nenciu, 2004]\label{T:2}
Given $\beta>0$, let $\alpha_k$, $0\leq k\leq 2n-2$, be distributed
as follows
\begin{equation}\label{DistAR}
\alpha_k \sim \begin{cases} B(\tfrac{2n-k-2}{4}\beta + a +
1,\tfrac{2n-k-2}{4}\beta + b + 1)    &
\text{$k$ even,} \\
B(\tfrac{2n-k-3}{4}\beta + a+b+2,\tfrac{2n-k-1}{4}\beta)        &
\text{$k$ odd.}
\end{cases}
\end{equation}
Let $\alpha_{2n-1}=\alpha_{-1}=-1$ and define
\begin{align}
b_{k+1}   &= (1-\alpha_{2k-1})\alpha_{2k} -
(1+\alpha_{2k-1})\alpha_{2k-2}          \label{E:BofG}\\
a_{k+1}   &= \big\{
(1-\alpha_{2k-1})(1-\alpha_{2k}^2)(1+\alpha_{2k+1}) \big\}^{1/2}
\label{E:AofG}
\end{align}
for $0\leq k \leq n-1$; then the eigenvalues of the tri-diagonal
matrix
$$
J = \begin{bmatrix}
 b_1 & a_1  &       &      \\
 a_1 & b_2  &\ddots &      \\
     &\ddots&\ddots & {\!a_{n-1}\!}     \\
     &      &{\!a_{n-1}\!}&  b_n
\end{bmatrix}
$$
are distributed according to the the Jacobi ensemble \eqref{E:JE}.
\end{theorem}

Let us note that one of the essential ideas behind the proof of this theorem
is to construct the distributions on Verblunsky coefficients and use the Geronimus
relations \eqref{AB:Geron} to transfer them to the Jacobi setting.

%%%%%%%%%%%%%%%%%%%%%%%%%%%%%%%%%%%%%%%%%%%%%%%%%%%%%%%%%%%%%%%%%%%%%%%%%%%%%%%%%%%%%
\section{The Ablowitz-Ladik system: Structure}
%%%%%%%%%%%%%%%%%%%%%%%%%%%%%%%%%%%%%%%%%%%%%%%%%%%%%%%%%%%%%%%%%%%%%%%%%%%%%%%%%%%%%

By analogy with the connection between the Toda lattice and Jacobi matrices,
we investigate the existence and properties of an integrable system
related to orthogonal polynomials on the unit circle and CMV matrices.
The main evolution of the system is defocusing Ablowitz-Ladik (also
known as the integrable discrete nonlinear Schr\"odinger equation). An excellent elementary introduction into
integrable systems is given in \cite{Deift}; \cite{Bab} is a detailed presentation of
many aspects of the subject. For the more advanced notions that we shall use
(especially regarding Lie and Poisson-Lie structures) we refer to \cite{IntSysII}.

The celebrated Toda lattice is the central example of a discrete integrable Hamiltonian system.
It models a 1-dimensional chain of particles with exponential nearest neighbor
interactions and it was introduced by Morikazu Toda, \cite{Toda}, to explain the findings of the
famous 1955 Los Alamos computer experiment of Fermi, Pasta, and Ulam, \cite{FPU}.
Complete integrability of the system was proved by Flaschka in 1974, \cite{Flaschka},
by introducing a change of variables that allowed him to set the system in Lax pair form.
Indeed, if Flaschka's variables $a_k$ and $b_k$ are set to be the entries of a Jacobi
matrix, $J$, then the evolution of the system is given by the equation
$$
\dot J=[J,P],
$$
with
$$
P=J_+-J_-=
\begin{bmatrix}
 0   & a_1 &      &       & \\
-a_1 & 0   & a_2  &       & \\
    & -a_2 & 0    &\ddots & \\
    &      &\ddots&\ddots & \\
\end{bmatrix}.
$$
Thus the Toda lattice is intrinsically connected to Jacobi matrices and orthogonal polynomials on the real line $\IR$.

In \cite[Section 11.11]{Simon2}, Barry Simon and the author found that the integrable system naturally related
to the theory of orthogonal polynomials on the unit
circle is the defocusing Ablowitz-Ladik system. It
was introduced in 1975--76 by Ablowitz and Ladik, \cite{AL1,AL2}, as a
space-discretization of the cubic nonlinear Schr\"odinger equation (NLS).
It reads
\begin{equation}\label{ALE1}
-i\dot\beta_n=\rho_n^2 (\beta_{n+1}+\beta_{n-1})-2\beta_n,
\end{equation}
where $\beta=\{\beta_n\}\subset\mathbb{D}$ is a sequence of
complex numbers inside the unit disk and
$$
\rho_n^2=1-|\beta_n|^2.
$$
The analogy with the continuous NLS becomes transparent if we
rewrite \eqref{ALE1} as
$$
-i\dot\beta_n=\beta_{n+1}-2\beta_n+\beta_{n-1}
-|\beta_n|^2(\beta_{n+1}+\beta_{n-1}).
$$
From this point onwards, $\dot f$ denotes the time derivative of the function $f$.

If we change variables to $\alpha_k(t)=e^{2it}\beta_k(t)$, this system becomes
\begin{equation}\label{ALeqn}
-i \dot\alpha_k = \rho_k^2 ( \alpha_{k+1} + \alpha_{k-1} ),
\end{equation}
which is a little simpler.  If we then choose $\alpha_{-1}$ and $\alpha_{n-1}$ to lie
on the unit circle, then they do not move and we obtain a finite system of ODEs for
$\alpha_k$, $0\leq k \leq n-2$, which is the specific case we treat.

Nenciu and Simon, \cite[Section11.11]{Simon2}, found that
if periodic Verblunsky coefficients are allowed to evolve according to Hamiltonians
in the Ablowitz-Ladik hierarchy, then their movement takes place on tori which are isospectral with respect to the CMV matrix. The
symplectic form they introduced extends naturally to the case of finite Verblunsky coefficients,
which we present here. In this setting it is given by:
\begin{align}
\{f,g\}_\text{AL} &= \sum_{j=0}^{n-2} \rho_j^2 \left[\frac{\partial
f}{\partial u_j}\frac{\partial g}{\partial v_j}-
 \frac{\partial f}{\partial v_j}\frac{\partial g}{\partial u_j}\right]
\label{PB2:uv}\\
&=2i \sum_{j=0}^{n-2} \rho_j^2 \left[\frac{\partial f}{\partial
\bar\alpha_j}\frac{\partial g}{\partial \alpha_j}-
 \frac{\partial f}{\partial \alpha_j}\frac{\partial g}{\partial
 \bar\alpha_j}\right]\label{PB2:alpha}
\end{align}
for any smooth functions $f,g:\ID^{n-1}\rightarrow\IR$, where $\alpha_j=u_j+i v_j$ for all $0\leq j\leq n-2$, and, as usual,
$$
\frac{\partial}{\partial\alpha}=\frac12
\left(\frac{\partial}{\partial u}-i\frac{\partial}{\partial
v}\right) \quad\text{and}\quad \frac{\partial}{\partial\bar\alpha}
=\frac12 \left(\frac{\partial}{\partial u}+i\frac{\partial}{\partial
v}\right).
$$
For clarity, we will call this bracket the Ablowitz-Ladik
bracket.  Note that this differs by a factor of two from that used in \cite{Simon2} and \cite{N}; we will
adjust the results accordingly.

On the set of Verblunsky coefficients $(\alpha_0,\ldots,\alpha_{n-2})\in \ID^{n-1}$ (with fixed $\alpha_{n-1}\in S^1$)
we consider the Hamiltonians $\Re(K_m)$ and $\Im(K_m)$, where
$$
K_m=\tfrac1m \tr(\C^m)
$$
for $m\geq1$. These form the Ablowitz-Ladik hierarchy. Indeed,
the evolution of the Verblunsky coefficients under
the flow generated by $\Re(K_1)$ is the Ablowitz-Ladik evolution:
$$
\{\alpha_j,\Re(K_1)\}_\text{AL}=\dot\alpha_j=i\rho_j^2(\alpha_{j-1}+\alpha_{j+1}).
$$

The evolution equations for the CMV matrices under these flows in the Ablowitz-Ladik bracket
were found by Nenciu~\cite{N}:

\begin{theorem}[Nenciu, 2005]\label{T:LP}
The Lax pairs for the $m^{\rm{th}}$ Hamiltonian of the finite
defocusing Ablowitz-Ladik system are given by
\begin{equation}\label{LPK}
\{\C,\Re(K_{m})\}_{AL}=[\C,i\C^{m}_+ + i(\C^{m}_+)^*]
\end{equation}
and
\begin{equation}\label{LPKbar}
\{\C,\Im(K_{m})\}_{AL}=[\C,\C^{m}_+ - (\C^{m}_+)^*]
\end{equation}
for all $m\geq1$, where for a matrix $\mathcal A$ we set
$\mathcal{A}_+$ as the matrix with entries
$$
(\mathcal{A}_+)_{jk}=\left\{%
\begin{array}{ll}
    \mathcal{A}_{jk}, & \quad\hbox{if}\,\,j<k;\\
    \tfrac12\mathcal{A}_{jj}, & \quad\hbox{if}\,\,j=k;\\
    0, & \quad\hbox{if}\,\,j>k.\\
\end{array}%
\right.
$$
\end{theorem}

The system of nonlinear differential-difference equations
\begin{equation}\label{SchurFlows}
\dot\al_n=(1-\al_n^2)(\al_{n+1}-\al_{n-1}),\quad
\{\al_j\}\subset(-1,1),
\end{equation}
is known as the discrete modified KdV equation (see \cite{AL1} and
\cite{G}) or the equation of the Schur flows (see \cite{FG}). This
system's main interest lies in its connection to the Toda and
Volterra (or Kac-van Moerbeke) lattices. More precisely, the Schur flows appear in the work of Ammar and
Gragg \cite{AG} as an evolution equation on Verblunsky
coefficients obtained by transferring the Toda equation via the
Geronimus relations from the $a$'s and $b$'s. As such, it is an
evolution on real $\al$'s. One can then relax this condition and
think of it as an evolution on complex coefficients, which preserves
reality (that is, if the $\al$'s are real at time 0, then they
remain real for all times).

As it turns out, this description applies to ``half" of the
Ablowitz-Ladik flows, the ones generated by $\Im(K_m)$ for all
$n\geq1$. Indeed, the evolution on the Verblunsky coefficients under
these Hamiltonians is equivalent to the Lax pairs
$$
\{\C,\Im (K_m)\}_{AL}=[\C,(\C^m)_+ -((\C^m)_+)^*],
$$
which preserve reality of the $\al$'s. So it is natural to ask ourselves what are
the corresponding evolutions of the $a$'s and $b$'s.
Moreover, we recover \eqref{SchurFlows} as the
evolution generated by $-\Im(K_1)$ under the Ablowitz-Ladik bracket
\eqref{PB2:uv}, \eqref{PB2:alpha}. In fact, in the finite case, all the evolutions induced by
$\Im(K_n)$ become, via the Geronimus relations, simple combinations
of the evolutions in the Toda hierarchy (this can be deduced from the
evolution of the corresponding spectral measures from \cite{KN05}).

There exists an abstract way of defining Poisson brackets on Lie algebras.
In particular, if $\g$ denotes the (associative) algebra of $n\times n$ complex
matrices, then the algebra structure gives rise to a natural Lie algebra
structure:
$$
  [B,C] = BC-CB.
$$
Of course, this also results from viewing $\g$ as the Lie algebra of
$GL(n;\Cmplx)$.
As a vector space, $\g=\LL\oplus\A$, where
$$
\A = \{A : A = - A^\dagger \},
$$
which is the Lie algebra of the group $\U(n)$ of $n\times n$ unitary
matrices, and
$$
\LL = \{A\in\g : L_{i,j} = 0 \text{ for $i>j$ and } L_{i,i}\in\Reals
\}
$$
which is the Lie algebra of the group $\LLL(n)$ of $n\times n$ lower
triangular matrices with positive diagonal entries. We will write
$\pi_{\A}$ and $\pi_\LL$ for the projections into these summands.

This vector-space splitting of $\g$ permits us to give it a second
Lie algebra structure. First we define the classical R-matrix $R:\g\to\g$ by
$R=\pi_\LL-\pi_{\A}$.
The second Lie bracket can then be written as either
\begin{equation}\label{E:DefnRBracket}
\begin{aligned}{}
[X,Y]_{\RR} = \tfrac12 [R(X),Y] + \tfrac12 [X,R(Y)]     \quad&\text{$\forall$ $X,Y\in\g$, or} \\
[L+A,L'+A']_{\RR} = [L,L'] - [A,A']    \quad&\text{$\forall$
$L,L'\in\LL$, and $A,A'\in\A$.}
\end{aligned}
\end{equation}
Using this R-matrix, one can define several Poisson brackets on the Lie algebra $\g$.
The Lie-Poisson (or Kirillov) bracket associated to the Lie algebra $(\g,[\,,]_\RR)$ is linear,
and it is relevant to Jacobi matrices: under the embedding $J\mapsto iJ$, the manifolds of Jacobi matrices
with fixed trace are symplectic leaves.
In our case, we focus on the Gelfand-Dikij (or quadratic) bracket, denoted here by $\{\cdot,\cdot\}_{GD}$.
We refer to \cite{IntSysII} for the definitions and proofs of the general results of the theory.

\begin{theorem}[Killip, Nenciu, 2005]\label{T:SympLeaf}
The manifold of CMV matrices with fixed determinant form a symplectic leaf in the Poisson manifold
$(\g,\{\cdot,\cdot\}_{GD})$.
\end{theorem}

The proof of this theorem relies on the fact a unitary matrix having the same
shape as a CMV matrix is exactly a CMV matrix. Moreover, one can show that
the set of CMV matrices with fixed determinant is an orbit under the action of a certain group of
dressing transformations (see Theorem~4.7 of \cite{KN05}). In a concurrent paper, \cite{Li}, Luen-Chau Li has
independently derived this result.
In a sense, his approach is the reverse of ours: he begins by investigating the action of certain
dressing transformations, while we arrive at their existence only after studying the problem by
other means.

As is natural in the theory of Poisson-Lie algebras, we focus on Hamiltonians given by
$\phi(B)=\Im\tr f(B)$, where $f$ is a polynomial. The general theory guarantees that the
evolution equations they generate will be expressible as Lax pairs. In our case
these Lax pairs at a CMV matrix are the same as those obtained from
Theorem~\ref{T:LP} by taking the appropriate linear combinations. Given this result, the following
is not surprising:
\begin{theorem}[Killip, Nenciu, 2005]\label{T:Equal}
For any $0\leq k,l\leq n-2$ the Gelfand-Dikij brackets of the
Verblunsky coefficients are given by
\begin{equation}\label{E:BrEqual}
\{\alpha_k,\alpha_l\}_{GD}=0\quad\text{and}\quad\{\alpha_k,\bar\alpha_l\}_{GD}=-2i\delta_{kl}\rho_k^2.
\end{equation}
\end{theorem}

That is, the restriction of the Gelfand-Dikij bracket to CMV matrices agrees with
the Ablowitz-Ladik bracket when written in the coordinates given
by the Verblunsky coefficients. From now on we denote both brackets by $\{\cdot,\cdot\}$.

%%%%%%%%%%%%%%%%%%%%%%%%%%%%%%%%%%%%%%%%%%%%%%%%%%%%%%%%%%%%%%%%%%%%%%%%%%%%%%%%%%%%%%%%%%%%%%%%%%%%%%%%%%%%%
\section{The Ablowitz-Ladik system: Asymptotics}
%%%%%%%%%%%%%%%%%%%%%%%%%%%%%%%%%%%%%%%%%%%%%%%%%%%%%%%%%%%%%%%%%%%%%%%%%%%%%%%%%%%%%%%%%%%%%%%%%%%%%%%%%

Shortly after the discovery of the Lax pair representation for the
Toda lattice, Moser, \cite{Moser}, gave a complete solution for the
finite system. Specifically, he discovered the angle
variables associated to the actions of H\'enon \cite{Henon} and
Flaschka \cite{Flaschka}.  In addition, he studied the long-time
asymptotics of the system and determined the scattering map.

The identification of the Ablowitz-Ladik bracket as an abstract
Poisson-Lie bracket allows one to follow the same path in the unitary setting.
More precisely, we begin by investigating the evolution under a Hamiltonian
of the Ablowitz-Ladik hierarchy of the spectral measure
$\mu=\sum \mu_j\delta_{z_j}$ associated to a CMV matrix $\C$ and the vector
$e_1=[1,0,\ldots,0]^T$:
\begin{theorem}[Killip, Nenciu, 2005]\label{C:mudot}
Under the flow generated by $\phi(B)=\Im\tr\{f(B)\}$,
\begin{equation}\label{E:CCC}
\partial_t \,(\log[\mu_j(t)]) = \{\phi, \log[\mu_j] \} = F(e^{i\theta_j}) - \sum_{l=1}^n F(e^{i\theta_l})\mu_l(t),
\end{equation}
where $F(z) = 2 \Re z f'(z)$.  Consequently,
\begin{equation}\label{E:MuEvol}
    \mu_j (t)  = \frac{\exp[F(e^{i\theta_j})\,t]\,\mu_j(0)}{\sum \exp[ F(e^{i\theta_l})\,t ]\,\mu_l(0)  }
\end{equation}
and for any $j,l\in\{1,\ldots,n-1\}$,
\begin{equation}\label{E:CanCom}
    \{ \theta_l , \tfrac12 \log[\mu_j / \mu_n ] \} = \delta_{jl}.
\end{equation}
\end{theorem}

A special case of this result (for Schur flows) has already appeared, \cite{MukNak}.  The
approach used there was to begin with a special case of
\eqref{E:MuEvol} and determine the induced evolution on the
Verblunsky coefficients.

In defining $\mu_k$ and $z_k$, we can choose any ordering we please;
however, there is a particular condition on this choice that
simplifies the formulae below.  Namely, we require that
\begin{equation}
 \lambda_1 \geq \lambda_2 \geq \cdots \geq \lambda_n,
\end{equation}
where we use the shorthand $\lambda_k = F(z_k)=2\Re[ z_k f'(z_k)]$.
Of course generically, the ordering will be strict. Note that by the continuity of $F$
this labeling is well-defined on an open set which is invariant under the flow of $\phi$.

In what follows we will assume that we are indeed in a situation where the ordering described above is
strict. This simplifies the formulae, but is not necessary; the interested reader can find the
full results in \cite{KN05}.

Under the flow generated by the Hamiltonian $\phi$, the masses have the following asymptotics
\begin{equation}\label{AsymptMu}
\log[\mu_k(t)] = -(\lambda_1 - \lambda_k)t + \log\biggl[\frac{\mu_k}{\mu_1}\biggr] +O(e^{-at})
\end{equation}
as $t\rightarrow\infty$. (The quantities for which we do not specify time
dependence are assumed to be evaluated to $t=0$.)
In particular $\mu_1(t)\to1$ and, if $k>1$, then $\mu_k(t)\to0$ exponentially fast.
We now turn to the asymptotics of the Verblunsky
coefficients:
Fix $1\leq k \leq n-1$. As all $\lambda_j$ are distinct,
\begin{equation}\label{E:AsympB}
\alpha_{k-1}(t) = (-1)^{k-1} \bar z_1\cdots \bar z_k \left[ 1 +
    \xi_{k-1} e^{-(\lambda_{k}-\lambda_{k+1})t} + O( e^{-\gamma t} )  \right]
\end{equation}
where
\begin{equation}
\xi_{k-1} = (z_k \bar z_{k+1} - 1) \frac{\mu_{k+1}}{\mu_{k}}
    \prod_{l=1}^{k-1}\left|\frac{z_{k+1}-z_l}{z_{k}-z_l}\right|^2
\end{equation}
and $\gamma>(\lambda_{k}-\lambda_{k+1})>0$.

In other words, if all $\lambda_j$ are distinct, then viewed as a curve in the
disk, $\alpha_{k-1}(t)$ approaches the boundary in a fixed
non-tangential direction.  This simply amounts to the statement that
$\xi_{k-1}$ is non-zero and $\arg(\xi_{k-1})=\arg(z_k \bar z_{k+1} -
1)$ belongs to $(-\pi/2,\pi/2)$. Let us also note that the
asymptotics of $\rho_{k-1}$ are easily deduced from
\eqref{E:AsympB}:
\begin{align*}
  \rho_{k-1}^2 (t) &= -2\Re(\xi_{k-1}) e^{-(\lambda_{k}-\lambda_{k+1})t} + O( e^{-\gamma t} )  \\
  &= | z_{k+1}-z_k |^2 \frac{\mu_{k+1}}{\mu_{k}}
    \prod_{l=1}^{k-1}\left|\frac{z_{k+1}-z_l}{z_{k}-z_l}\right|^2 e^{-(\lambda_{k}-\lambda_{k+1})t} + O( e^{-\gamma t} ).
\end{align*}
This shows that the factors $\mathcal{L}(t)$ and $\mathcal{M}(t)$ of the CMV matrix $\C(t)$
diagonalize as $t\to\infty$ and hence so does $\C(t)$.  Moreover, the eigenvalues are ordered
by the corresponding value of $F(z)$. This is a well known phenomenon for the Toda lattice.

Let us note here that when the $\lambda_j$ are not all distinct, $\C(t)$ converges to a direct sum of CMV matrices and
their adjoints.  Specifically, if
$$
  \lambda_{k-1} > \lambda_k=\cdots=\lambda_{k+m} > \lambda_{k+m+1},
$$
then $\alpha_{k-1},\ldots,\alpha_{k+m-2}$ do not approach the
boundary and $\C(\infty)$ has a non-trivial block of size $m$
beginning at row/column $k$.  If $k$ is odd, this will be a CMV
matrix; if $k$ is even, it will the adjoint of a CMV matrix.
While this phenomenon cannot occur for the Toda lattice, it can
occur for Hamiltonians in the same hierarchy.  Some examples of
non-diagonalization are discussed on page~389 of \cite{DLT_JFA85}.

Finally, our understanding of the asymptotic behaviour of the relevant
quantities allows us to find canonical coordinates for the system. We know from
the Lax pairs that the eigenvalues Poisson commute: $\{\theta_j,\theta_k\}=0$,
for all $1\leq j,k\leq n$, where $\theta_j$ is the argument of the eigenvalue $z_j$. Moreover,
the evolution of the spectral measure given in Theorem~\ref{C:mudot} implies that
$$
\{ \theta_l , \tfrac12 \log[\mu_j / \mu_n ] \} = \delta_{jl}.
$$
It would be natural to imagine that
\begin{equation*}
  \theta_1,\ldots,\theta_{n-1},\tfrac12\log[\mu_1/\mu_n],\ldots, \tfrac12 \log[\mu_{n-1}/\mu_n]
\end{equation*}
are canonical coordinates. This is the case for the Toda lattice (as can be concluded
from \cite{DLNT_CPAM86}). The same does not happen here; instead, the following holds:
\begin{theorem}[Killip, Nenciu, 2005]\label{P:OhDear}
For any labelling of the eigenvalues,
\begin{equation}\label{E:OhDear}
\bigl\{ \log[\mu_2/\mu_1] , \log[\mu_3/\mu_1] \bigr\} = 2\cot\bigl(\tfrac{\theta_1-\theta_2}2\bigr)
+ 2\cot\bigl(\tfrac{\theta_2-\theta_3}2\bigr)
+ 2\cot\bigl(\tfrac{\theta_3-\theta_1}2\bigr)
\end{equation}
in the Gelfand--Dikij (or Ablowitz--Ladik) bracket.
\end{theorem}

While the previous results concerning asymptotics for the Ablowitz-Ladik system are
proved using the abstract (Gelfand-Dikij) form of the bracket, we were unable to do the same here.
Instead note that $\{\log[\mu_j/\mu_l],\log[\mu_k/\mu_l]\}$ is constant under the flow generated
by a Hamiltonian $\phi$ defined as above,
and so one can work in a regime where the masses are already exponentially ordered. We
then use perturbation theory to obtain more precise asymptotics for $\log[\mu_j/\mu_1]$ for
$j=2,3$ in this regime, and plug these into the explicit (Ablowitz-Ladik) bracket to obtain the result.
It would be interesting to find a proof using only the Gelfand-Dikij bracket, and to better understand
the appearance of the cotangent function in formula \eqref{E:OhDear}.

In an earlier paper, \cite{KN}, Rowan Killip and the author considered the map
\begin{equation}\label{E:COV}
(\theta_1,\mu_1,\ldots,\theta_{n-1},\mu_{n-1},\theta_n) \mapsto (u_0,v_0,\ldots,u_{n-1},v_{n-2},\phi)
\end{equation}
where $\alpha_k=u_k+iv_k$ and $\alpha_{n-1}=e^{i\phi}$. We found that:
\begin{prop}\label{P:Jac}
The Jacobian of the change of variables \eqref{E:COV} is given by
$$
\det\left[\frac{\partial (u_0,v_0,\ldots,\phi)}{\partial(\theta_1,\mu_1,\ldots,\theta_n)}\right] =
  - 2^{1-n} \frac{\rho_0^2\cdots\rho_{n-2}^2}{\mu_1\cdots\mu_n}.
$$
\end{prop}
The analogous result for Jacobi matrices is due to Dumitriu and Edelman, \cite{DumE}, which served as our guide.
Both of these proofs are probabilistic in nature.

We also asked for a simpler, more direct derivation of the Jacobian. A recent preprint of Forrester
and Rains, \cite{ForrR}, gives a very direct, computational solution.  Percy Deift showed us a derivation of the
Jacobi matrix result using the symplectic structure naturally associated to the Toda lattice.
The same can be done in the circular case (see \cite[Corollary~8.6]{KN05}).  The key idea is the following:
As we can write the underlying symplectic form in either set of variables, we can view \eqref{E:COV}
as a symplectomorphism between two concrete symplectic manifolds.  In particular, it must preserve
the Liouville volume, which allows one to prove Proposition~\ref{P:Jac}.

\medskip

\noindent \textbf{Acknowledgements:} The author wishes to
thank the organizers of the CRM 2005 Meeting on Random Matrices,
Random Processes and Integrable Systems for the opportunity to attend and
to speak there. She is also grateful to Percy Deift and Barry Simon for
their support and helpful discussions. Much of the author's recent work on
different aspects of CMV matrices is joint with Rowan Killip.

%%%%%%%%%%%%%%%%%%%%%%%%%%%%%%%%%%%%%%%%%%%%%%%%%%%%%%%%%%%%%%%%%%%%%%%%%%%%%%%%%%%%%%%%%

\end{document}